\documentstyle[pra,preprint,aps,epsf]{revtex}
\tighten
\draft
\begin{document}
\begin{center}
{\large\bf Unconditional Security Of Quantum Key Distribution
Over Arbitrarily Long Distances\footnote{This reprint version contains the
same material as the one published in {\it Science} {\bf 283}, 2050--2056
(1999). We also include the refereed supplementary notes (as in
{\tt http://www.sciencemag.org/feature/data/984035.shl}) explicitly in the
appendix for easy reference.}} \\
~~~\\
Hoi-Kwong Lo$^1$\footnote{e-mail: hkl@hplb.hpl.hp.com} and
H. F. Chau,$^2$\footnote{e-mail: hfchau@hkusua.hku.hk}\\
~~~\\
$^1$ Hewlett-Packard Laboratories, Filton Road, Stoke Gifford, Bristol
 BS34 8QZ, UK.\\
$^2$ Department of Physics, University of Hong Kong, Pokfulam Road,
 Hong Kong, China\\
~~\\
(\today)
\end{center}
\begin{abstract}
Quantum key distribution is widely thought to offer
unconditional security in communication between two users.
Unfortunately,
a widely accepted proof of its security in the presence of
source, device and channel noises has been missing.
This long-standing
problem is solved here by showing that, given fault-tolerant
quantum computers, quantum key distribution over an  arbitrarily long
distance of a realistic noisy channel can be made unconditionally secure.
The proof is reduced from a  noisy quantum scheme to a noiseless quantum scheme and then
from a noiseless quantum scheme to a noiseless classical scheme,
which can then be tackled by classical probability theory.
\end{abstract}

\newpage
\widetext

The art of secure communication --- cryptography --- has a long history.
Before two parties can communicate securely,
they often must
share  a secret random string of numbers (a key) for
encryption and decryption. The secrecy
of the message depends on the secrecy of the key.
A problem in conventional cryptography is the key distribution problem: 
In classical physics, there is nothing to prevent an
eavesdropper from monitoring the key distribution channel passively,
without being caught by the legitimate users.

Quantum key distribution (QKD) (1-5)
has been proposed as a solution to the problem.
The quantum no-cloning theorem states that it
is impossible to make an exact copy of an unknown quantum state
(6). Thus, it is generally thought
that eavesdropping on a quantum channel will almost surely
produce detectable disturbances.
The two users can, therefore, use part of their
quantum signals to test for eavesdropping.
Only when the error rates are acceptable
will they use the quantum signals to
generate a key. Thus, the two users (commonly called Alice and Bob)
have the confidence that if an eavesdropper (commonly called Eve)
has a nonnegligible amount of information
on the final key, she will almost surely be caught,
even if she
has infinite
computing power and access to a quantum computer.
Incidentally,
several recent experiments have demonstrated
the feasibility of QKD over tens of kilometers (7).

``The most important question in quantum cryptography is
to determine how secure it really
is '' (8, p.16). QKD is widely claimed to provide perfect security.
However, this viewpoint has been under renewed scrutiny
for two reasons. First, contrary to
well-known claims of
unconditional security (9), a class
of other quantum cryptographic schemes,
including so-called
quantum bit commitment and
quantum one-out-of-two oblivious transfer,
has recently been shown to be insecure
(10) . Cheaters can defeat these schemes by
a subtle application of
the well-known Einstein-Podolsky-Rosen (EPR) paradox (11)
and by delaying their measurements.
These ``no-go'' theorems not only
shattered the long-standing belief in the security of those schemes,
but they also undermined the confidence in QKD itself.
Second, a convincing and rigorous proof of the security
of QKD has been missing despite extensive investigations
(12-15). Thus, the foundation of quantum cryptography has been shaky. Here, we
solve this long-standing
problem by proving that, given quantum computers, QKD can be made
unconditionally secure
over arbitrarily long distances.

A rigorous proof of the security of a QKD scheme requires the explicit
construction of a procedure such that,
whenever Eve's strategy has a nonnegligible probability of
passing the verification test by Alice and Bob,
her information on the final key will be exponentially small (16-17).
This procedure must be secure and
efficient, even when Alice and Bob use
imperfect sources and devices and share a noisy quantum channel.

Most analyses on the security of  QKD have
dealt with single-particle eavesdropping strategies (12),
with immediate or delayed measurements, as well as
the so-called collective attacks (13),
in which Eve brings each signal particle into interaction with a
separate probe system but then,
after hearing the public discussion between Alice and Bob,
measures all probes together.
Security against the most general type of attack, the so-called joint attack,
has been investigated by Deutsch {\it et al.} and also by Mayers. The discussion by
Deutsch {\it et al.} was restricted to the special case of
perfect devices (14). It introduced the concept of
quantum privacy
amplification, based on a process called entanglement purification,
which was studied by Bennett, DiVincenzo, Smolin and Wootters (BDSW) (18).
Earlier versions of Mayers' proof (15) have not been
accepted as definitive. His most recent version of the proof 
is more detailed and
complex (19). He proposes a proof of security of the Bennett and Brassard (BB84) (2) scheme
against joint attacks in the presence of detector and channel noise
but with an ideal trusted single-photon source.
Our current work and work by Mayers (19) are contemporaneous
and independent. They differ
greatly in their premises, methods and consequences:
(i) Mayers' work deals with the standard BB84 QKD protocol (2)
for preparation,
transmission, and measurements of nonorthogonal states. His approach
does not require Alice and Bob to have a quantum computer, although
Eve may have one.
In contrast, our proof applies to a new QKD protocol,
involving fault-tolerant sharing and purification of so-called
EPR pairs, and requires
that Alice and Bob have quantum computers.
(ii) Mayers' work (19) assumes an ideal
single-photon or EPR-pair source, thus disallowing a beam-splitter attack.
[A testing procedure for an allegedly ideal
EPR-pair source from an untrusted vendor has recently been suggested
by Mayers and Yao (20).] In contrast,
our work allows the reception of
untrusted imperfect quantum signals from the channel
(21).
(iii) Our proof and protocol allow QKD to be securely extended over
arbitrarily large distances through a chain of insecure relay stations.
A similar extension of BB84 in Mayers' proposed proof would require secure
relay stations, to
which Eve does not have access.
(iv) Our proof is conceptually simpler.
(v) Our techniques have widespread applications outside QKD.

Why is a proof of security of QKD so difficult? In a joint attack,
Eve treats the whole sequence of quantum
signals as a single entity. She couples this entity with her probe and then
unitarily evolves the combined system.
She forwards a subsystem to Bob and  keeps the remaining subsystem for
eavesdropping purposes. Eve can use any unitary transformation
she likes, and yet, a secure QKD scheme must defeat all of them.
Moreover, Eve may attempt to mask her presence by attributing the errors
caused by her eavesdropping attack to normal transmission noise. Furthermore,
because the particles are now generally entangled with each other,
a naive application of  classical probability theory may
lead to fallacies [See the EPR paradox (11)].

Despite these apparent difficulties,
 we show that it is possible to distinguish a malicious
Eve from noise. Moreover, it is
possible to use classical probability theory to establish the
security of QKD.

{\bf Techniques and importance of results.}
Assuming that users have access to quantum computers, we
show the security of QKD by a reduction in two steps.
The central theme of the first step is to reduce the
noisy quantum scheme (imperfect devices, noisy channels, storage errors,
and so forth)
to a noiseless quantum scheme. We do this by
combining the ideas of ``quantum repeaters''
(22,23) and fault-tolerant quantum computation (FTQC)
(24,25). Although these are
existing ideas in the field, we make the nontrivial observation that
they can be combined and applied to QKD
to distinguish noise from a malicious Eve.
In particular, we note that
knowing the error syndrome does not help an eavesdropper.
Therefore, we can give an eavesdropper full control of the quantum
repeater stations without compromising security.

Even in a noiseless quantum scheme, Alice and Bob are required
to verify that the particles are untampered by Eve.
Things will be easy if one can apply classical arguments to
solve this quantum problem at hand. However, as illustrated
by the EPR paradox, naive classical arguments often lead to fallacies.
The most important technical contribution of this paper is our second theme ---
reducing the noiseless
quantum verification scheme to a classical one.
Finally, we establish the security of the classical verification scheme by
classical probability theory. The security of the quantum scheme then
follows.

The use of classical arguments in our quantum
problem allows us to simplify our discussion greatly.
We emphasize that the validity of this usage is highly paradoxical.
Classical arguments work in our quantum problem
because all the observables $O_i$'s under consideration are diagonal
with respect to a single basis
$\cal B$.
In more detail, let us consider the observable $M$, which represents a
complete von Neumann measurement along the same basis $\cal B$.
Because all of the $O_i$'s under consideration
are diagonal with respect to $\cal B$,
$M$ commutes with all the observables $O_i$'s.
Therefore, the measurement 
$M$ along basis $\cal B$ will in
no way change the outcome of the subsequent 
$O_i$'s. Without any loss of generality,
one can imagine that such a complete von Neumann measurement $M$
is always performed before the measurement of subsequent 
$O_i$'s.
In other words, the
initial state of the quantum system is simply a
classical mixture of
eigenstates of $M$, and hence, classical
arguments carry over to the quantum case.
We remark that the  $O_i$'s that we 
consider
are coarse-grained observables (observables with
degenerate eigenvalues)
rather than fine-grained observables (observables with
nondegenerate eigenvalues).

{\bf Quantum-computational protocols.}
The execution of our secure QKD scheme requires large-scale quantum computers
for both error correction and verification.
Building such computers is a technological feat that is far beyond our current
technology. However, all existing QKD security analyses require some
idealization also. In an actual
experimental implementation of polarization-coding BB84 (a standard
``prepare-and-measure'' P/M scheme) over a
substantial distance (say 40km) of a lossy quantum channel
using coherent states, Eve may,
in principle, break the
system completely by a generalized beam-splitting attack (26).
This is so
even when the bit error rate  of the quantum signals is
strictly zero.

Quantum-computational protocols like ours  are
worthy of analysis for several reasons. First, unlike the
usual P/M schemes, they extend the range of
secure QKD to arbitrarily long distances even with
insecure quantum repeaters. Second, when implemented over a noisy channel
without repeaters, it is conceivable that they can tolerate a higher noise
level than a
standard P/M scheme. Third, a proof of security and the tradeoff between
noise and key rate are much easier than those for P/M schemes.
Indeed, our scheme provides a conceptually simple
and rigorous proof of the security
of QKD without  the full complexity of
a P/M scheme.

{\bf EPR pairs.}
Before we report our QKD scheme in detail, let us first recapitulate the
usefulness of an EPR pair, that is, a singlet state ${ 1 \over \sqrt{2}} (|\uparrow
\downarrow\rangle - | \downarrow\uparrow\rangle)$ of
a pair of quantum bits (qubits) (27) in QKD.
If two members of an EPR pair are measured along
any common axis, each member will give a random outcome,
and yet, the outcomes of the two members will always be
antiparallel.
This spooky action at a distance defies any simple classical explanation
and is at the core of EPR paradox (11).

Now, suppose two distant users share $R$ EPR pairs. Then,
the random outcome of measurement along a common axis
generates an $R$-bit key between them.
The laws of quantum physics assure that the key
is truly random and that Eve cannot have any information
on its value. Indeed, the two lemmas in supplementary material 
(available at www.sciencemag.org/feature/data/984035.shl ) [that is,
Supplementary Note 2 in this reprint version]
show that, to generate an almost perfectly secure $R$-bit key, Alice
and Bob only need to share $R$ EPR pairs of almost perfect fidelity (28).
Therefore, all we need for secure QKD is a way for Alice and
Bob to share EPR pairs and to verify that, indeed, they are EPR pairs. 
We  focus on these EPR distribution and verification
problems. There are two issues that Alice and Bob
have to address: noise and Eve.

{\bf Reduction to a noiseless scheme.}
One can classify errors in an
EPR-pair distribution process into four types.
First, the quantum communication channel between
Alice and Bob is generally noisy. Second, the EPR source may be
imperfect in itself. Third,
errors may occur during the storage
of quantum information. Fourth, errors may also occur during computation;
because elementary gates and measuring devices for quantum computation
are generally imperfect, gate errors and measurement errors may arise.

The last three types of errors can be fixed by recently developed
quantum error correction (29)
and fault-tolerant quantum computation (FTQC) (24,25) techniques.
In particular, there is a ``threshold result'' in
FTQC: Assuming an independent noise model and that
the error rates for each primitive computational
gate and for each time step of storage are smaller
than some positive threshold values, one can perform arbitrarily long quantum computations
with an arbitrarily high fidelity (25).
The essence of FTQC is to defeat errors by
encoding quantum states in quantum error-correcting codes (QECC) and then
performing quantum computation on the encoded states.

{\bf Quantum repeaters.}
We must consider the first type of noise ---
channel noise.
If the quantum communication channel is very noisy
(for example, it is very long), we cannot apply the
threshold result in
FTQC to
combat quantum communication errors. Fortunately, the idea of
quantum repeaters has been proposed
as a much more efficient way of correcting quantum communication
errors (23). The idea is summarized as follows .

Given impure EPR pairs shared between two distant observers,
they can apply local operations and classical communication to
distill a smaller number of higher fidelity EPR pairs in a
procedure known as entanglement purification (18).
However, for distances much longer than the coherence length of a
noisy quantum communication channel, the probability that a quantum state
will remain error-free
is exponentially small.
Therefore, the fidelity of transmission is
so low that standard purification methods are not applicable.

Quantum repeaters are essentially simple quantum computers installed
throughout a quantum communication channel. They are used to divide
the channel into shorter segments, which are then purified separately,
before they are connected. The number and locations of quantum repeaters are
chosen so that it is possible to
create EPR pairs with sufficiently high fidelity between the two
ends of each segment.
After creating EPR pairs that are shared between the two ends,
one applies entanglement purification by using
quantum repeaters. This will, at the cost of discarding some pairs,
increase the fidelity of the remaining pairs.
Afterwards, EPR pairs shared between various segments are connected together by
``quantum teleportation'' (30).
Indeed, a highly efficient procedure involving a sequence of entanglement
purification and teleportation has been devised that allows
the reliable sharing of EPR pairs between two arbitrarily distant
locations (23).

Three important remarks are in order. First, with
two-way classical communication, quantum repeaters can greatly
improve the yield of distillation (18,23)
over the standard fault-tolerant circuits. Second,
even highly imperfect quantum repeaters can do the job very well:
It has been argued (23) that an error rate 
in the percent level is readily tolerable.
Third,
a strength of our approach is that, assuming perfectly reliable local
quantum operations by Alice and Bob, one can actually calculate the threshold
value for tolerable noise between two adjacent quantum repeaters.
For example, in the case of a depolarizing channel, it is
known that a fidelity of
$1/2$ is the threshold value (18).

With quantum repeaters and FTQC,
the usual threshold result can be extended to
distributed quantum computation over a realistic noisy quantum channel
(31). In other words,  any distributed quantum
algorithm, including the
EPR-pairs distribution process, that works in the
noiseless case can always be
extended to the noisy case.

{\bf Error syndrome contains no useful information for an eavesdropper.}
We make the most generous assumption that Eve completely controls the
quantum repeaters and the quantum communication channel. Alice and Bob need
only trust their own
quantum computers and authenticated classical messages from each other.
There is a subtlety for us to address.
If Eve follows the correct procedure, she will not
be caught. However, she does learn about the error syndrome
(that is, the pattern of measurement results) generated during
a FTQC, which allows the error-correction apparatus to correct a
corrupted state to a former value.
So, the question is Can the error syndrome tell her anything useful
about the state?
The answer is``no" because of the following.
 Mathematically, each of Alice and Bob's state
can be written as a tensor product of the logical qubits
(which actually contain the quantum information) and the ancillary
qubits (which contain the error syndromes) (32); that is, the wave function
$| \Psi \rangle$ can be written as $ \sum_{i,j} c_{ij} |a_i \rangle_L \otimes |e_j \rangle_A $,
where subscripts $L$, $A$, $i$ and 
$j$ represent the logical qubits, the ancillary qubits, the logical state (in some orthonormal basis), and
the error syndrome, respectively, and $c_{ij}$'s are some complex coefficients. (In reality, Alice
and Bob's system is generally entangled with Eve's probe. However, this does not
change the essential point of our argument.) Because of FTQC,
although the state of the ancillary qubits evolves
unpredictably
over time, the state of the logical qubits will, with
a very high fidelity [$1 - O (e^{-\ell})$ for some arbitrarily chosen
$\ell> 0$ (33)], follow the desired
computation and remain unaffected by the errors.
As long as the gate error rate and storage error
rate are sufficiently small, the subsequent verification
and key generation steps can be thought of as being performed solely
on the logical qubits.
In other words, the ancillary qubits decouple from the verification step.
Accordingly, 
we shall ignore the ancillary qubits and focus only on the logical
qubits.

If there is no appreciable eavesdropping, the logical
qubits will represent the desired state.
Of course, an honest Eve can learn as much as she likes about the
error syndrome. The general theory of QECC
tells us that the error syndrome contains absolutely no information
about the encoded quantum state (29).

In summary, we have reduced the proof of security
of our noisy QKD (or EPR delivery) scheme to that of
a noiseless one. Now, we focus on the noiseless scheme.

{\bf The goal of verification.}
To make sure that there is no substantial eavesdropping,
Alice and Bob must verify that the state of the logical qubits
is, indeed, that of $N$ EPR pairs.
Without any loss of generality, we can allow Eve to not merely  act on the $N$ EPR
pairs while they are being shared but to  actually  prepare them in an arbitrary
state of her choosing and then give them to Alice and Bob (14).
She claims that they are perfect EPR pairs.
Alice and Bob will be happy to sacrifice a small number $m$ of those pairs
to verify Eve's claim. If any one of the
$m$ tested pairs fails the test, then all the $N$ pairs are discarded.
However, if all the $m$ pairs pass the test, the remaining $N-m$
pairs will be accepted as singlets and used to generate the key.
The goal of the verification is for Alice and Bob to make sure that
Eve has a very small probability of cheating successfully. By cheating
successfully, we mean that the $m$ tested pairs
pass the verification test and yet the
remaining $N-m$ pairs, if given a yes or no test of being
$N-m$ singlets, will give ``no'' as an answer (34).

The security of our
quantum verification scheme will automatically
guarantee the security of the corresponding QKD scheme [refer to (28) for an explicit bound on Eve's information].
 We will now consider the security of our quantum verification scheme.

Essentially, what Alice and Bob are trying to do
is to distinguish singlets from triplets.
Although there is no way for Alice and Bob to do so with certainty using only
local operations and classical communication, they can do so with a very high probability.

The goal of a quantum verification scheme is to verify that the state
of the $N$ pairs is, in fact, $N$ singlet. A 
direct testing of a random subset of
EPR pairs requires an exponential amount of
resources in term of the security parameter $k$, where the probability for
Eve to cheat successfully is, at most,  $e^{-k}$.
Direct testing of a random subset is,
therefore, not an efficient verification scheme. To understand this point,
suppose Eve cheats by inserting a single nonsinglet among the $N$ pairs
and only $m$ random pairs are tested by Alice and Bob.
There is a probability $N-m \over N$ for this nonsinglet to remain
untested. Consequently, Eve has at least a probability $N-m \over N$
of cheating successfully. To prevent this from happening,
it is necessary for $m$ to be of order $N$.
Even when $m$ equals $N-1$, the probability for Eve to cheat
successfully is still at least $1/N$. For this to be exponentially
small in $k$,  the number of photons transmitted $N$, must be exponentially
large in $k$.

A much more efficient way of verifying a
quantum state exists.
It is due to the
random-hashing idea by BDSW (18).
(BDSW proposed it for error correction, but here we use it for
verification.)
It is in the same spirit of a classical random-hashing scheme which
we will now describe.

{\bf A classical verification scheme.} Imagine a game in which Eve
locks an $N$-bit string $x$ in a
box and Alice and Bob are allowed to ask a small number $m < N$ of ``fair''
questions about it,
which Eve must answer truthfully.
A fair question is a yes or no question whose answer is "yes" for
exactly half of all $N$-bit strings. Thus, Is the
first bit 1? and Are the first and third bits equal? are
fair questions, but Are all the bits 1's? is unfair.
If all of Eve's answers are consistent with the assumption
that the string $x$ is
all 1's,
Alice and Bob must ``accept'' the string.
Otherwise, Alice and Bob must ``reject'' the string.
Finally, Eve opens the box and shows Alice and Bob the string to
prove that she has answered faithfully.
Eve wins if
the string is not all 1's
and yet Alice and Bob have accepted the string.
[Alice and Bob win if
the string is not all 1's and they have
rejected the string. 
If the string is, in fact, all 1's, the game is a draw.]

If Alice and Bob ask only single-digit questions of the form
Is the $k$th bit a 1? then Eve has a good chance of winning by
choosing a string with a single $0$ at a random location.
However, if Alice and Bob instead ask Eve about the
parities of random subsets of the bits, they quite likely catch any string that is not all 1s.

For example,
if the unknown string is $x= 1101$ and Alice and
Bob choose a subset consisting of the second and third
bits (this can be represented conveniently by
an index string $s = 0110$),
the parity $x \cdot s$ is $1$. This test reveals that $x$ is not
all 1's, since an all-1's four-bit string would have had parity $0$ on this
subset.
More generally, the parity of a subset $s$ of the bits
in a string $x$ is the inner product,
or modulo-2 sum of the bit-wise AND of strings $x$ and $s$,
and it is  denoted by $x \cdot s$.
[In this example,  $x \cdot s = 1  \cdot 0 + 1  \cdot 1
+ 0  \cdot 1 + 1 \cdot 0 = 1 \pmod 2$.]
The probability that two different strings give the same answer for
$m$ iterations of random parity check is no more than $2^{-m}$ (18).
Thus, by checking only a few subset parities (say $20$), Alice and
Bob can reduce their chance of accepting an $x$ that is not all 1s to less
than one in a million.

 Eve must not know the index strings
beforehand. Otherwise, she could always cheat successfully,
in a similar way as a smuggler who knows beforehand which of the several
bags a customs inspector will open in an airport.
Indeed, because the string has $N$ bits and there are only $m$ constraints
(generated by $m$ rounds of parity
verification), there are clearly exponentially many ( namely, $2^{N-m}$),
strings that will pass the test. However,
because Eve does not know the index strings beforehand
and because the index strings are chosen randomly, Eve
effectively has to put her bet on a single string without prior knowledge.
We see from the last paragraph that
any string $x \not= 11 \cdots 1$ chosen by Eve has only an
exponentially small
probability ($2^{-m}$) of passing the verification test.

{\bf Our quantum verification scheme.} Now,
we construct an efficient quantum verification scheme that is similar to
the classical verification scheme that we have just described.
Consider the so-called Bell basis, $\Psi^{\pm} $ and $\Phi^{\pm}$, where
\begin{equation}
\Psi^{\pm}= { 1 \over \sqrt{2}}
( | \uparrow \downarrow \rangle \pm | \downarrow \uparrow \rangle)
\end{equation}
and
\begin{equation}
\Phi^{\pm}= { 1 \over \sqrt{2}}
( | \uparrow \uparrow \rangle \pm | \downarrow \downarrow \rangle).
\end{equation}
With the convention in (18),
Bell basis vectors are represented by two classical bits
\begin{eqnarray}
\Phi^+ & =& \tilde{0}\tilde{0}, \nonumber \\
\Psi^+ & =& \tilde{0}\tilde{1}, \nonumber \\
\Phi^- & =& \tilde{1}\tilde{0}, \nonumber \\
\Psi^- & =& \tilde{1}\tilde{1}.
\label{bellstate}
\end{eqnarray}
\noindent
(Because Bell basis vectors are maximally entangled, one
should never think of them as direct product states.)
A complete basis for $N$-ordered pairs of qubits (what
we shall call $N$-bell basis)
consists of products of Bell basis vectors,
each of which is described by a
$2N$-bit string.
In the absence of an
eavesdropper, Alice and Bob share
$N$ singlets, whose state is described by a $2N$-bit string of $\tilde{1}$'s,
$| \tilde{1}\tilde{1} \cdots
\tilde{1} \rangle$.

What happens when there is an eavesdropper?
Recall that
we allow Eve to not merely  act on the $N$ EPR
pairs while they are being shared, but to actually  prepare them in an arbitrary
state of her choosing and then give them to Alice and Bob.
The pairs may be entangled
among themselves as well as with a probe in Eve's hands (14).
A system described by any mixed state can be equivalently
described by a pure state of a larger
system consisting of the original system and an ancilla
(10,14).
As discussed by Deutsch {\it et al.} (14),
by considering the larger system instead,
we shall, without a loss of generality, consider
that Eve prepares a pure state
\begin{equation}
| u \rangle = \sum_{i_1, i_2 , \cdots, i_N} \sum_j
\alpha_{i_1, i_2 , \cdots, i_N,j} | i_1, i_2 , \cdots, i_N \rangle
\otimes | j \rangle ,
\label{two0}
\end{equation}
where $i_k$ denotes the state of the $k$th pair, which 
runs from $\tilde{0} \tilde{0}$ to $\tilde{1}\tilde{1}$, $\alpha_{i_1, i_2 , \cdots, i_N,j} $'s are
some complex coefficients, and the
$| j \rangle$ values
form an orthonormal basis
for the ancilla. Each state $| u \rangle$ represents a particular cheating
strategy chosen by Eve.

The goal of a quantum verification scheme is to verify that the string
describing the state of the $N$ pairs is, in fact,
all $\tilde{1}$'s. We now construct an efficient quantum verification scheme
based on the quantum random-hashing idea by
BDSW (18).
BDSW showed that
one can compute the parity of any subset of
the $2N$-bit string by using
local operations and classical communication only (35).
The parity is ``collected'' into a single destination pair;
it is determined by the outcomes of measurements
performed on that pair, which has to be discarded afterward.
More specifically, the
parity is found by noting whether the measurement outcomes on the two members
of the destination pair are parallel or antiparallel.

If Eve prepares a classical mixture of products of Bell states, it is not
too difficult to show that classical arguments apply
directly to the quantum
verification problem and Eve's probability of cheating successfully
is negligible [see (36)  for
details].

{\bf Why do classical arguments work for a quantum problem?}
However, 
instead of preparing a classical mixture of products
of Bell states, in the most general eavesdropping strategy as shown in Eq.4, Eve
prepares a general state, which is entangled with her probe.
The big question
is Can Eve prepare a more general state to enhance her
probability of cheating successfully?
The crux of our paper is the following claim:
If Eve prepares a general state to cheat in
the BDSW
random-hashing verification scheme, her
probability of cheating successfully will be 
exactly the same as in the situation when
she premeasures that state along the $N$-Bell basis before handing
it over to Alice and Bob.
In other words,
a general state offers no advantage  over a mixture of
products of Bell states. With this quantum to classical reduction
result, (36) applies to any eavesdropping strategy.
This proves that Eve's probability of cheating successfully is negligible
and our QKD scheme is secure against all possible
attacks.

{\bf Proof of our claim.} Consider the following observables on a state
$| u \rangle$ of $N$ pairs of qubits shared between Alice and Bob.
We define these observables by their actions on the
$2^{2N}$ $N$-Bell states, which form a complete basis. Let $W$,
defined by $W |w \rangle =
w | w \rangle$, be the observable that gives the $2N$-bit string representing
the state $w$ in BDSW notation. For any index string $s$, let $Q_s$, defined by
$Q_s |w \rangle = ( s \cdot w ) |w \rangle$, be the observable
that gives the parity of the subset $s$ of the bits. Finally, let
$R = | \tilde{1}\tilde{1} \cdots \tilde{1} \rangle
\langle \tilde{1}\tilde{1} \cdots \tilde{1} |$ be the projector onto a state
of $N$ singlets. All the above
operators refer to a single basis (namely, the $N$-Bell basis).
Because all the observables ($R$, $W$ and $Q_s$) are
simultaneously diagonalizable with respect to the $N$-Bell basis,
$R$ and all the $Q_s$ values commute with $W$.
Therefore, neither the
value of $R$ nor any of the $Q_s$ values are affected by a prior measurement of $W$. In other
words, for any state $| u \rangle$ that Eve might have supplied, neither the
sequence of subset parities measured in the verification stage nor the
result of the final hypothetical measurement of $R$ would have
been affected if
Eve had pre-measured $| u \rangle$ in Bell basis (that is, made a measurement
of $W$) before handing the state to Alice and Bob.
 Incidentally, the fact  that
a premeasurement does not change the outcomes of some subsequent
measurements  is highly reminiscent of  work by Griffiths
and Niu (37).

{\bf Subtleties in our proof.}
The following example illustrates the computation of parities and
the subtleties involved.
Suppose Alice and Bob share three pairs of qubits. With the procedure
specified in BDSW, the computation of the parity of
the first subset (for example, $s_1= 001101$) can be
done by the circuit diagram shown in Fig.~1A.
The parity is collected into a single
pair and is determined by whether that pair gives a parallel or
antiparallel outcome when both members are measured along the $z$ axis.

The computation itself, up to phases, performs a permutation on the
space of all $2N$-bit strings. After the computation, the measured pair
is dropped from consideration, and only two pairs remain.
The computation of the parity of the second subset (for example, $s_2= 1001$) by the
BDSW procedure is shown in Fig.~1B.
After the computation, another pair is measured and
dropped from consideration.
Therefore, only a single pair out of the original three is left
after the computation of the two parities (for $s_1$ and $s_2$).
A simple unitary description (10) of the overall computation is that
it maps, up to phases, the state
$| \tilde{1} \tilde{1} \tilde{1} \tilde{1} \tilde{1} \tilde{1} \rangle$ to
$| \tilde{1} \tilde{0} \tilde{1} \tilde{1} \tilde{1} \tilde{1} \rangle$.
Suppose also that, on passing the verification test, Alice
and Bob generate their secret key by measuring the remaining pair
along the $z$ axis, with an ``up'' for Alice's result meaning ``0'' and
a ``down'' meaning ``1''.

A number of subtleties deserve careful
discussions. First, as in the classical case,
the choice of subsets can be announced
only after Alice and Bob receive all of their
quantum particles. So long as Eve does not know the subsets beforehand,
her probability of cheating successfully is exponentially
small (see supplementary material, available at www. 
sciencemag.org/feature/data/984035.shl) [that is, Supplementary Note 4 in this
reprint version]. 

Second, during the computation of the parities of subsets,
the state of the $N$ pairs of qubits is transformed by a unitary
transformation $U_{s_1, s_2, \cdots, s_m}$, which
depends on the subsets $s_i$.
But would that unitary transformation $U_{s_1, s_2, \cdots, s_m}$
somehow spoil our reduction argument from a quantum to classical verification?
Fortunately, the answer is ``no".
Despite the apparent complexity of the parity computation
procedure, the bottom-line answer that Alice and Bob obtain is simply
the parities (that is, the eigenvalues of the operators $Q_s$'s of their choice).
Therefore, the verification test proves that, for any
general cheating strategy by Eve
that passes the test with a probability of at least $2^{-r}$,
the conditional fidelity of the state  before the parity computation
as $N$ singlets, $|\tilde{1} \tilde{1} \cdots  \tilde{1} \rangle$,
is $1 - O (2^{-(m-r)})$.
Consequently, the state after the
parity computation will, with the same fidelity, be
$U_{s_1, s_2, \cdots, s_m} |\tilde{1} \tilde{1} \cdots  \tilde{1} \rangle$.

Third, in our quantum verification procedure, Alice and Bob have to
disclose all their measurement outcomes in a public
channel. For each measured pair,
there are four possible outcomes, ``$\uparrow \downarrow$'',
``$\downarrow  \uparrow$'', ``$\uparrow \uparrow$''
and ``$\downarrow \downarrow$'', thus resulting in two bits of information.
This is more than the one-bit (0 or 1) parity information.
Now, the question is
Can Eve somehow benefit from this additional information?
The answer is ``no" (this discussion is available at  www. 
sciencemag.org/feature/data/984035.shl) [that is, Supplementary Note 5 in this
reprint version]. 
Finally, the issue of a quantum Trojan horse attack is addressed in
(21).
This completes our proof of security of QKD.


{\bf Discussion.}
An important idea behind our quantum to classical reduction is that
a quantum mechanical experiment has a classical interpretation
whenever observables that refer to only one basis (the $N$-Bell basis
in our case) are
considered.  The fine-grained
measurement operators by Alice
and Bob along the three random bases do not commute with the
Bell-basis projection operators. However, Alice and Bob base their
decision on whether to accept the alleged singlets not on those
fine-grained measurement results  but on the coarse-grained
(parallel or antiparallel)
ones. Those coarse-grained operators all commute with a
complete von Neumann measurement along
the Bell basis (38).

Our quantum to classical reduction technique is a powerful
tool of widespread applications.
It guarantees that one can apply standard results in the classical world
(such as probability theory and statistics theory)
to the original quantum problem without
leading to fallacies.
In effect, this means the extension of classical statistical theory
to quantum mechanics, resulting in a quantum statistical theory.
To illustrate this point, we give
two other examples of applications of our
quantum to classical reduction result . (i) Suppose two
distant observers share $N$ pairs of qubits, and estimate the
number of singlets in those $N$ pairs.
By the number of singlets, we mean the expected number of
``yes'' answers if a singlet or triplet measurement was
made on each pair individually.  (ii) Under the assumption
that signal carries are perfect single photons, put a probabilistic
bound on an
eavesdropper's information in BB84 as a function of the error rates of the
sampled photons. (These examples are discussed at www. 
sciencemag.org/feature/data/984035.shl.) [That is, Supplementary Note 6 in this
reprint version.]

The second example gives us a quantitative statement on
the trade-off between information gain and
disturbance (39). This is
a strong result to
a notoriously difficult problem because (i) the bound applies not merely to a
strategy in which Eve couples a probe to each signal particle but
to any information extraction strategy that is consistent with
quantum mechanics and (ii) the bound can be derived by a
random sampling of a small subset. In other words, a concrete
experimental random-sampling procedure (rather than
an abstract mathematical equation with little physical meaning)
is presented here (40).

Finally, let us return to QKD itself.
Although we have  focused on the
case when Alice and Bob receive allegedly good EPR pairs from Eve,
our proof of security of QKD also
applies  to the case when Alice sends qubits (rather than
halves of EPR pairs) to Bob.
Consider the following situation.
Alice prepares $N$ EPR pairs in her laboratory.
She then chooses the subsets for parity determination beforehand
and performs all the computations
and measurements on her halves of the $N$ EPR pairs
in her own laboratory before sending out the other halves to Bob.
After Alice's measurements,
the subsystem that she sends to Bob is in a pure state;
that is, qubits rather than halves of EPR pairs are sent to Bob.
However, because Alice's operation is local, it
must commute with Eve's eavesdropping operator.
Therefore, this qubit-based scheme must be as secure as the
original EPR-based scheme. (Just as in the EPR-based case,
it is of the utmost importance for Alice to withhold information on 
the choice of subsets for the parity determination
until Bob acknowledges the receipt of quantum transmission.
Otherwise, Eve can cheat easily.)

\newpage
\begin{figure}
 \vspace{2in}
 \hspace{0.2in}
 \epsfxsize=16cm
 \epsfbox{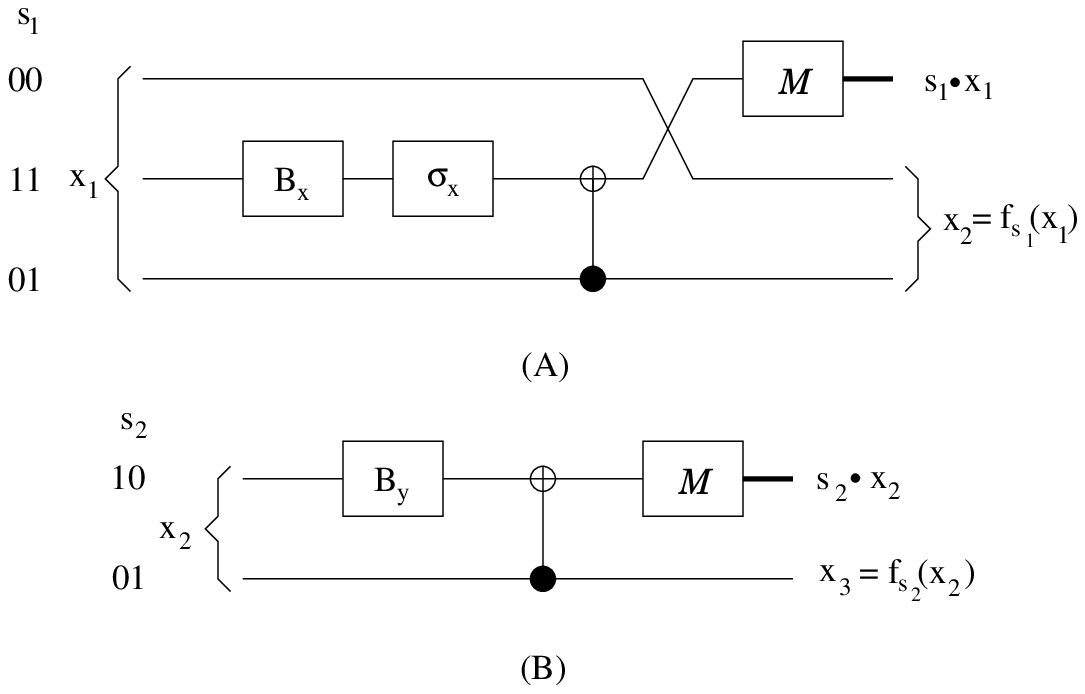}
 \caption{A sample one-way hashing protocol used to determine $s_1 \cdot x_1$
  and $s_2 \cdot x_2$ for an unknown three-Bell state. Following the
  convention of (18), $B_x$ and
  $B_y$ denote bilateral rotation of $\pi/2$ along the $x$ and $y$ axis,
  respectively; $\sigma_x$ denotes a unilateral rotation of $\pi$ along the
  $x$ axis; and the symbol $\bullet\!\mbox{---}\!\oplus$ denotes a bilateral
  controlled NOT operation. $\mathit{M}$ denotes a bilateral measurement.
 }
 \label{F:1}
\end{figure}
\newpage
{\bf Appendix: Refereed Supplementary Materials}
\par\medskip\noindent
{\em Supplementary Note 1:}
\par\medskip
The beamsplitter attack
has been discussed by C. H. Bennett, F. Bessette, G. Brassard, L. Salvail, and
J. Smolin in Ref.~\cite{ssingle}.
Here we show that, even in the case of zero bit error rate (BER) in
BB84, a generalized version of beamsplitter attack can break BB84
completely provided that
the loss due to the quantum channel between Alice and Bob is sufficiently
large. For ease of discussion, we consider
a generalized beamsplitter attack, which has been
brought to our attention by N. L\"{u}tkenhaus.
Eve measures the photon number observable $N$ of each
pulse. If $N$ is less than two, she keeps the signal herself
and Bob receives nothing.
If $N$ is two or larger, she uses an ideal beamsplitter to
take one photon out. She then resends the remaining photons to Bob via a
{\it superior} channel so as to ensure that Bob still gets the same bit
rate.
[This is possible provided that the loss of the quantum channel
between Alice and Bob is large enough.]
Eve stores her photons and waits for the announcement of the
polarization bases. As far as polarization is concerned,
Eve has an exact copy of
Bob's signals. Therefore, by measuring her own photons
along the correct bases in BB84,
she learns the polarizations
of Bob's received photons
and hence the key.

\par\medskip\noindent
{\em Supplementary Note 2:}

\par\medskip
{\noindent \it Lemma~1}:  (High fidelity implies low entropy)
If $ \langle R \mbox{~singlets} | \rho | R  \mbox{~singlets} \rangle >
1-\delta$ where $\delta \ll 1$, then
the von Neumann entropy $S(\rho) < - ( 1 - \delta) \log_2 ( 1 -
\delta) -  \delta \log_2 {\delta \over  (2^{2R} -1)}$.
\par\medskip\noindent
{\noindent \it Proof:} If $ \langle R \mbox{~singlets} | \rho | R
\mbox{~singlets}
\rangle > 1- \delta$, then the largest eigenvalue of the density matrix $\rho$
must be larger than $ 1- \delta$, the entropy of $\rho$ is, therefore, bounded
above by that of $\rho_0 = {\rm diag}~\{1- \delta, {\delta \over  (2^{2R}
-1)}, {\delta \over  (2^{2R} -1)}, \cdots ,{\delta \over  (2^{2R} -1)} \}$.
That is, $\rho_0$ is diagonal with a large entry $1- \delta$ and with the
remaining
probability $\delta$ equally distributed between the remaining $2^{2R} -1$
possibilities.
\hfill Q.E.D.
\par\medskip
{\noindent \it Lemma~2}: (Entropy
is a bound to mutual information.) Given any
pure state $\phi_{AB}$
of a system consisting of two subsystems
$A$ and $B$, and any generalized measurements $X$ and $Y$
on $A$ and $B$ respectively, the entropy of each subsystem $S( \rho_A)$
(where $\rho_A = {\rm Tr}_B | \phi_{AB} \rangle \langle  \phi_{AB} |$)
is an upper bound to the amount of mutual information between $X$ and $Y$.
\par\medskip
{\noindent \it Proof:} This is a corollary to Holevo's theorem
(See A. S. Holevo,
 {\it Probl.\ Inf.\ Transm.\ (USSR)\ } {\bf 9}, 117 (1973).)
\hfill Q.E.D.
\par\noindent\medskip

\par\medskip\noindent
{\em Supplementary Note 3:}

\par\medskip
For instance, one can simply choose a block code of length $k$ (say
$k=7$) 
that can correct a single error
to encode a single qubit repeatedly. In the first level of
this concatenated coding scheme, the qubit is mapped into $k$ qubits.
In the second level of the coding, each of the $k$ qubits is
individually encoded into $k$ qubits, resulting in $k^2$ qubits
altogether, so on and so forth. Let us assume that the errors
for various elementary gates are
independent and of order $\epsilon$.
After the first level of encoding, all single errors can be corrected.
Therefore, the effective error rate is of the order $\epsilon^2$.
More generally, the effective error rate of the $(L+1)$-th level is
related to that of the $L$-th level by $\epsilon^{(L+1)} \sim
( \epsilon^{(L)})^2$. Consequently, one expects that
$\epsilon^{(L)} \sim
\epsilon_0 ( { \epsilon \over \epsilon_0 })^{2^L} $
where $\epsilon_0$ is some threshold value of the error rate.
(See Eq.\ (36) of Preskill in Ref.~\cite{ftqc}.)
Therefore, the error rate becomes exponentially suppressed
whenever a level is added.

\par\medskip\noindent
{\em Supplementary Note 4:}
\par\medskip
On the contrary, if Eve were to know the subsets
beforehand, then similar to our discussion on classical random hashing,
Eve could alway cheat successfully:
In the above example, suppose Eve is told beforehand
the subsets $s_1$ and $s_2$ and that
Alice and Bob will generate their key by measuring their pair along $z$-axis.
Then, Eve can cheat by finding out which cheating final state
will pass the verification test and then following its evolution
backwards in time to work out the initial state:
Suppose Eve would like the value of the key to be ``0''.
She observes that the final state
$|\tilde{1}\tilde{0}\tilde{1}\tilde{1}\rangle \otimes |\uparrow_z \downarrow_z
\rangle = \frac{1}{\sqrt{2}}
 |\tilde{1}\tilde{0}\tilde{1}\tilde{1}\rangle\otimes
(|\tilde{1}\tilde{1}\rangle + |\tilde{0}\tilde{1}\rangle)$ will
achieve her goal. Evolving it backwards in time, Eve finds
that she needs to prepare the initial state
$\frac{1}{\sqrt{2}} |\tilde{1}\tilde{1}\tilde{1}\tilde{1}\rangle\otimes
(|\tilde{1}\tilde{1}\rangle - |\tilde{0}\tilde{1}\rangle)$.

\par\medskip\noindent
{\em Supplementary Note 5:}
\par\medskip
The key point to note is that, starting with an $N$-singlet state
$|\tilde{1} \tilde{1} \cdots  \tilde{1} \rangle$, the parity
computation will simply evolve it into another $N$-Bell state, up to a
phase \cite{BDSW}. Notice that the $N$ pairs in such a final
state are {\it not} entangled with one another.
Therefore, if we consider the untested $N-m$ pairs, they
should be in a {\it pure} state. In fact, they will be
in an $(N-m)$-Bell state,
described by the density matrix,
${\rm Tr}_{\rm tested}
U_{s_1, s_2, \cdots, s_m} |\tilde{1} \tilde{1} \cdots  \tilde{1} \rangle
\langle \tilde{1} \tilde{1} \cdots  \tilde{1}
| U_{s_1, s_2, \cdots, s_m}^{\dagger}$.

Recall that, for any {\it effective} eavesdropping
strategy (that is, one that passes the verification test
with a probability $\geq 2^{-r}$), the fidelity of the initial
state as $N$-singlets is very close to $1$.
By unitarity, the fidelity of the
final state as
$U_{s_1, s_2, \cdots, s_m} |\tilde{1} \tilde{1} \cdots  \tilde{1} \rangle$
is also close to $1$. Since fidelity does not decrease under
tracing [See R. Jozsa, {\it J.\ Mod.\ Opt.\ } {\bf 41}, 2315 (1994) for this
property.],
the fidelity of the subsystem of the untested pairs as
${\rm Tr}_{\rm tested}
U_{s_1, s_2, \cdots, s_m} |\tilde{1} \tilde{1} \cdots  \tilde{1} \rangle
\langle \tilde{1} \tilde{1} \cdots  \tilde{1} | U_{s_1, s_2, \cdots,
s_m}^{\dagger}$ is very close to $1$.

As ${\rm Tr}_{\rm tested}
U_{s_1, s_2, \cdots, s_m} |\tilde{1} \tilde{1} \cdots  \tilde{1} \rangle
\langle \tilde{1} \tilde{1} \cdots  \tilde{1} | U_{s_1, s_2, \cdots,
s_m}^{\dagger}$
is a pure state, we can apply the argument in note~\cite{Note1}
to prove that the von Neumann entropy of those $N-m$
untested pairs is very close to $0$ and, hence, that the mutual information
between those $N-m$ pairs and
the external universe (Eve plus tested pairs plus anything else) is
exponentially small. Since the tested pairs are just part of the
external universe, measurements on them do not really help.

\par\medskip\noindent
{\em Supplementary Note 6:}
\par\medskip
Here we give the key arguments for
the reduction result in
the two examples rigorously, but leave out
the irrelevant detailed calculations based on
classical statistical theory.
In the first example, consider $N$ pairs of qubits
shared between Alice and Bob. Those pairs can be entangled
with each other and also with the external universe,
for example, an ancilla prepared by Eve. Suppose Alice and Bob would like to
estimate the number of singlets. i.e., the
expected number of ``yes'' answers if a singlet-or-not
measurement were performed on each pair individually.
This number can, in principle, be
determined by Bell measurements if they bring the two halves together.
However, Alice and Bob would like to perform local measurements
and classical communication only.
We argue that they can, nonetheless, estimate it accurately by the following
method. They {\it randomly} pick $m$ pairs and measure the two
members of each pair along an
axis chosen randomly from $x$, $y$ and $z$ axes.
Then they publicly announce their outcomes.

{\it Claim}: If $k$ of the outcomes are anti-parallel, then the estimated
fraction
of singlets in the $N$ pairs is $(3k-m)/2m$. Furthermore,
confidence levels can be deduced from classical statistical theory.

{\it Proof}: Consider, for each pair, the projection operators
$P^i_{\parallel}$ and $P^i_{{\rm anti}-\parallel}$
for the two coarse-grained outcomes (parallel and antiparallel) of
the measurement performed on the $i$th pair. It is straight-forward to express
these projection operators as linear combinations of projection operators along
a single basis, namely, $N$-Bell basis.
Let us consider the operator $M_B$ which represents the action of
a measurement along $N$-Bell basis.
Since $M_B$, $P^i_{\parallel}$ and $P^i_{{\rm anti}-\parallel}$
all refer to a single basis ($N$-Bell basis), they clearly commute with
each other. Thus, a pre-measurement $M_B$ by Eve along
$N$-Bell basis will in no way change the outcome for
$P^i_{\parallel}$ and $P^i_{\parallel}$. Therefore, we may
as well consider the case when such a pre-measurement is performed
and the problem is classical to begin with.

Random sampling is a powerful technique in classical statistical
theory. The Central Limit Theorem shows that the asymptotic
or large-sample distribution of the mean of a random sample from
{\it any} finite population with finite variance is normal.
See, for example, B. W. Lindgren, {\it Statistical Theory}
(3rd. ed., Collier Macmillan, London, 1976).
Therefore, for a sufficiently large $m$, we can simply use
a normal distribution to estimate the original mean
and establish the confidence level.

There is a minor subtlety.
For each pair,
only one of three random measurements is performed.
Since everything is now classical and is unrelated to the
key points of this paper,
we shall skip the details here and refer the readers to, for example,
W. G. Cochran, {\it Sampling Techniques} (3rd ed.,Wiley, New York, 1977),
chap.~5.
\hfill Q.E.D.

This result is profound because it allows one to apply
classical sampling theory and central
limit theorem to a quantum problem at hand.

For the second example, one can establish a rigorous probabilistic
bound on the eavesdropper's
mutual information $I_{\rm Eve}$ on the final key by random sampling
(i.e., by sampling
a random subset of photons, computing the error rates in the two bases
and then applying classical probability theory to estimate the real error
rates, etc). We assume that the signal carriers in
BB84 are perfect single
photons. Notice that Alice could use EPR pairs to prepare photons
in BB84. Let us
consider a pre-measurement $M_B$ along $N$-Bell basis by Eve
and the coarse-grained projectors
$P^i_{\parallel}$ and $P^i_{{\rm anti}-\parallel}$.
As before, they refer to a single basis and, thus, commute with
each other. Therefore,
a pre-measurement $M_B$ will in no way change
those coarse-grained outcomes.

Alice and Bob pick $m$ photons randomly from those that
are both transmitted and received along the {\em same} basis and
publicly compare their polarizations.
Alice and Bob can then
work out the ``typical subspace'' that is likely to give such
error rates and compute its dimensions. The logarithm of the
number of
dimensions of the typical subspace
will be a very good probabilistic
bound on the eavesdropper's information.
[The probability amplitude on the ``atypical subspace''
will also contribute to the entropy, but this
contribution can be made to be negligible in comparison.
In more detail, Suppose there are $N$ EPR pairs and the
squared amplitude on the atypical subspace is at most
$\epsilon$. Then the atypical space's contribution $S_a$
to the entropy is
no more than $S_a =
- \epsilon \log_2 { \epsilon \over
2^{2N}} \approx
 2N \epsilon $. By making sure that $\epsilon$ is
much less than $1$, $S_a$ is much less than $N$.
Since
the contribution of the typical space to the entropy is supposed to
be linear in $N$ in a noisy channel,
clearly $S_a$ is negligible in comparison.]


\begin{references}
\bibitem{Wiesner} The idea of quantum cryptography was first proposed
by S. Wiesner around 1970 but remained unpublished until 1983
[ S. Wiesner, {\it SIGACT News} {\bf 15} (no. 1), 78 (1983)].
Wiesner proposed quantum money and multiplexing channel (essentially
one-out-of-two oblivious transfer) but not QKD per se.
\bibitem{BB84} C. H. Bennett and G. Brassard, in {\it Proceedings of the IEEE
International Conference on Computers, Systems, and Signal  Processing}
(IEEE Press, New York, 1984), pp.175-179.
\bibitem{Ekert} A. K. Ekert, {\it\prl} {\bf 67}, 661 (1991).
\bibitem{book} H.-K. Lo, S. Popescu, and T. Spiller  {\it
Introduction to Quantum Computation and Information} (World Scientific,
 Singapore, 1998).
\bibitem{Loreview} For a review on quantum cryptography, see,
for example, H.-K. Lo, in (4), p.~76-119.
\bibitem{nocloning} W. K. Wootters and W. Zurek, {\it Nature} {\bf 299},
 802 (1982); D. Dieks, {\it Phys. Lett. A} {\bf 92}, 271 (1982).
\bibitem{QKD_Expt} P. D. Townsend, {\it Electron.\ Lett.\ } {\bf 30},
 809 (1994);
 A. Muller, H. Zbinden, N. Gisin, {\it Europhys.\ Lett.\ } {\bf 33},
 335 (1996); R. J. Hughes, G. G. Luther, G. L. Morgan, C. G. Peterson,
 C. Simmons, in {\it Advances in Cryptology: Proceedings of CRYPTO'96}, vol. 1109 of 
{\it Lecture Notes in Computer Science,} N. Kobiltz, Ed. (Springer-Verlag, Berlin, 1996),
 pp. 329-342. For a review, see, for example, H. Zbinden, in (4),
 p.~120-142.
\bibitem{25} G. Brassard and C. Cr\'{e}peau, {\it SIGACT News} {\bf 27} (no. 3),
13 (1996).
\bibitem{BCJL} G. Brassard, C. Cr\'{e}peau, R. Jozsa, D. Langlois,
 in {\it Proceedings of the 34th annual IEEE Symposium on the Foundation of
Computer Science} (IEEE Computer Society Press, Los Alamitos, CA, 1993),
pp. 362-371, and references cited therein.
\bibitem{bit_comm} For the impossibility of bit commitment, see the following:
 D. Mayers, Los Alamos e-Print archive (available at http://xxx.lanl.gov/abs/quant-ph/9603015);
  H.-K. Lo and H. F. Chau, {\it\prl} {\bf 78}, 3410 (1997);
 D. Mayers, {\it ibid.,}  p. 3414;
 H.-K. Lo and H. F. Chau, {\it Physica\ D}, {\bf 120}, 177 (1998).
 For the impossibility of other schemes, including one-out-of-two
oblivious transfer, see
 H.-K. Lo, {\it\pra} {\bf 56}, 1154 (1997).
 For a review, see, for example,
 H. F. Chau and H.-K. Lo, {\it Fort.\ de\ Phys.\ } {\bf 46}, 507 (1998)
 ; (5).


\bibitem{Bell} J. S. Bell, {\it Physics\ } {\bf 1}, 195 (1964); 
[reprinted in {\it Quantum Theory and Measurement,} J. A. Wheeler and W. H. Zurek, Eds.
(Princeton Univ. Press, Princeton, 1983), pp.403-408]; A. Einstein,
 B. Podolsky, N. Rosen, {\it Phys.\ Rev.\ } {\bf 47}, 777 (1935)
[reprinted in {\it Quantum Theory and Measurement,} J. A Wheeler and W. H. Zurek, Eds. 
(Princeton Univ. Press, Princeton, 1983), pp.~138-141].

\bibitem{ssingle} 
C. H. Bennett, F. Bessette, G. Brassard, L. Salvail,
 J. Smolin, {\it J.\ Cryptol.\ } {\bf 5}, 3 (1992);
 N. L{\rm \"u}tkenhaus, {\it\pra} {\bf 54}, 97 (1996);
 C. H. Bennett, T. Mor, J. A. Smolin, {\it ibid.,} p.  2675;
 C. A. Fuchs, N. Gisin, R. B. Griffiths, C.-S. Niu, A. Peres,
 {\it ibid.,} p. 1163 (1997); R. B. Griffiths and C.-S. Niu,
 {\it ibid.,}  p.1173; N. L{\rm \"u}tkenhaus
 and S. M. Barnett, Los Alamos e-Print archive (available at http://xxx.lanl.gov/abs/quant-ph/9711033).

\bibitem{scollect} E. Biham and T. Mor, {\it\prl} {\bf 78}, 2256 (1997);
 E. Biham, M. Boyer, G. Brassard, J. van de Graaf, T. Mor,
 Los Alamos e-Print archive (available at http://xxx.lanl.gov/abs/quant-ph/9801022).


\bibitem{sdeutsch} D. Deutsch {\it et al.}, {\it\prl} {\bf 77}, 2818 (1996); D. Deutsch {\it et al.}, {\it ibid.} 
{\bf 80}, 2022 (1998).
\bibitem{smayers} D. Mayers, in {\it Advances in Cryptology: Proceedings
 of CRYPTO'95}, vol. 963 of {\it Lecture Notes in Computer Science,} D. Coppersmith, Ed.  (Springer-Verlag,
 Berlin, 1995), pp. 124-135; in
 {\it Advances in Cryptology: Proceedings
 of CRYPTO'96}, vol. 1109 of {\it Lecture Notes in Computer Science,} N. Koblitz, Ed.  (Springer-Verlag,
 Berlin, 1996), pp. 343-357.
\bibitem{problem} In some applications, a nonnegligible
amount of information leakage to Eve can be disastrous.
The following example is due to J. Smolin (41).
If a
key is used in a so-called one-time pad to encrypt a president's
order which ends with the password for
launching a nuclear missile, an adversary who is
aware of the structure of the message
will, in principle, be able to steal the password.

\bibitem{negli} The goal of making Eve's expected information small conditional
only
on passing the test, is generally unattainable.
One must include the proviso:
 for any eavesdropping strategy with a nonnegligible chance of success.
\bibitem{BDSW} C. H. Bennett, D. P. DiVincenzo, J. A. Smolin, W. K.
 Wootters, {\it\pra} {\bf 54}, 3824 (1996).

\bibitem{Mayers} D. Mayers, Los Alamos e-Print archive 
(available at http://xxx.lanl.gov/abs/quant-ph/9802025),
version~4.



\bibitem{Mayao} D. Mayers and A. Yao, in {\it Proceedings of 39th Annual Symposium on
Foundations of Computer Science} (IEEE Computer Society Press, Los Alamitos, CA, 1998), 
pp. 509-515 [also available at
Los Alamos e-Print archive (available at http://xxx.lanl.gov/abs/quant-ph/9809039)].

\bibitem{remark} A big worry in cryptography is the Trojan horse attack.
Any untrusted material received from an open channel poses serious
security risks.
As J. Smolin has often remarked (41),
it is even conceivable that a robot is hidden in the
received material and that it pops out to find and disclose secrets
to adversaries. 
It is not just that the Trojan horse might leak information once it is in
Bob or Alice's laboratory. One might think that this problem could be eliminated 
by simply shielding the laboratory very 
well or that such shielding is , in fact, assumed anyway in cryptographic protocols.
The real problem is that the Trojan horse pretends to be real EPR pairs when Alice and Bob do their testing but
behaves differently when they generate key, thus causing them to leak the information themselves.
This worry is not unfounded because
it is notoriously difficult to prepare almost perfect EPR pairs.
[See (20) for a related discussion.]
Real quantum systems often contain other degrees of freedom
which are ignored in quantum computation.
One might wonder if
Eve could perform a quantum Trojan horse attack by hiding robots in
(the hidden Hilbert space dimensions of)
the quantum systems received by Alice and Bob.
This would certainly make a rigorous proof of security of
QKD based on imperfect sources impossible.
Our answer is the following proposition.
{\it Proposition~1.}As long as there is no security risk
for Alice and Bob to receive
untrusted classical messages, quantum Trojan horse attack can be foiled.
{ \it Remark.} Before we present our proof, notice that
the assumption that there is no security risk in receiving
classical messages is most reasonable
because Eve can always send classical messages to Alice and Bob in
a ``man-in-the-middle'' attack during a
classical authentication process.
If Alice and Bob could not afford to receive any untrusted classical
message, the whole enterprise of cryptography would be hopeless.
{ \it Proof:} Instead of receiving any untrusted
quantum system directly from an
open quantum channel, a user (say Bob) demands that the state of the system
must be converted
into classical messages via teleportation (30) right at his doorstep.
More concretely, Bob
prepares trusted EPR pairs in his laboratory
and sends one member of each pair
to his untrusted representative Robert, who is working
in an insecure area just outside his laboratory,
when the untrusted quantum data (potentially a Trojan horse)
is waiting. Robert teleports the nominal state of the
untrusted system (that is, the state in its nonclandestine variables)
into Bob's laboratory. In other words, Bob conveys
the untrusted quantum state into his laboratory by means of trusted
EPR pairs and untrusted classical messages.
Now assuming that there is no security risk in
receiving classical messages, Bob can safely receive those classical
messages and use them to reconstruct the
original quantum state. Teleportation
provides an exact counting of the effective dimensions of the
Hilbert space
because each qubit requires two classical bits to teleport.
Therefore, there is no hidden Hilbert space to worry about in the
reconstructed quantum system. This conclusion is valid even if the
original EPR pairs prepared by Bob do contain hidden dimensions.


\bibitem{relay1} S. J. van Enk, J. I. Cirac, P. Zoller, {\it\prl} {\bf 78},
 4293 (1997); J. Preskill, {\it Proc.\ R.\ Soc.\ London\ A\ } {\bf 454},
 385 (1998); J. I. Cirac, A. K. Ekert, S. F. Huelga, C. Macchiavello, 
 Los Alamos e-Print archive (available at http://xxx.lanl.gov/abs/quant-ph/9803017).
\bibitem{relay2}  H.-J. Briegel, W. D\"{u}r, S. J. van Enk,
 J. I. Cirac, P. Zoller, {\it Philos. Trans. R. Soc. London Ser. A } {\bf 356},
1713 (1998);
 H.-J. Briegel, W. D\"{u}r, J. I. Cirac, P. Zoller, {\it\prl}
 {\bf 81}, 5932 (1998);
 W. D\"{u}r,  H.-J. Briegel, J. I. Cirac, P. Zoller, {\it\pra} {\bf 59}, 169 (1999).

\bibitem{Shor} P. W. Shor, in {\it Proceedings of the 37th Symposium on Foundations of 
 Computer Science} (IEEE Computer Society Press, Los Alamitos, CA, 1996), pp. 56-65.
\bibitem{ftqc} A. Yu. Kitaev, {\it Russ. Math. Surv.} {\bf 52}, 1191 (1997);
 D. Aharonov and M. Ben-Or, in {\it Proceedings
 of the 29th Annual ACM Symposium on the Theory of Computing}, (ACM Press, New York, 1998),
 pp. 176-188;
 E. Knill, R. Laflamme, W. Zurek, {\it Science\ } {\bf 279}, 342 (1998);
 for a review, see, for example, J. Preskill, in (4), pp. 213-269.


\bibitem{attack} For instance, in the study of standard
P/M schemes such as BB84 (2), one often
assumes that the signal carriers are perfect single photons.
Unfortunately, producing almost perfect single-photon pulses is 
beyond current technology, and dim coherent light pulses
with
a Poisson distribution in the number of photons are often used instead.
The attenuation of an optical fiber is also large (say 0.35~dB/km),
and detector efficiencies are far from perfect. Therefore,
rather surprisingly, in an actual
experimental implementation of polarization-coding BB84 over a
significant distance (say 40km), Eve may,
in principle, break the
system by a generalized beamsplitting attack.
The key point is that, many of the signals contain more than
one photons and as such Eve is allowed to make copies
(details are available at http://xxx.lanl.gov/abs/quant-ph/984035.shl)
[that is, Supplementary Note 1 in this reprint version].
For short-distance applications, the relevance of such an attack
remains an important subject for future investigations.
In summary, standard theoretical security analyses on BB84 do
not apply to most real-life experimental systems to date.


\bibitem{qubits} A qubit  is simply a
 two-level quantum system.
It plays the role of a fundamental unit of quantum information,
 just like a bit in classical information.
\bibitem{Note1} Alice and Bob share $R$ EPR pairs and they
 generate a key by measuring these pairs along any common axis.
If the fidelity of their pairs is high (that is,
 $\langle R\mbox{~singlets} | \rho | R  \mbox{~singlets} \rangle >
1- 2^{-k}$ for a sufficiently large
$k$), then Eve's information on the final key will be bounded by
$2^{-c} + \mbox{O} (2^{-2k})$ where
$c = k -\log_2 ( 2R +k +  { 1 \over \log_e 2} ) $.
In other words, Eve's information (more precisely,
mutual information with the final key)
is exponentially small as a function of $k$.
This result follows directly from two lemmas (see
discussion, available at http://xxx.lanl.gov/abs/quant-ph/984035.shl)
[that is, Supplementary Note 2 in this reprint version]. 

\bibitem{qee} P. W. Shor, {\it\pra} {\bf 52}, 2493 (1995);
 A. M. Steane, {\it\prl} {\bf 77}, 793 (1996).

\bibitem{tele} C. H. Bennett {\it et al.}, {\it \prl} {\bf 70}, 1895 (1993).

\bibitem{Realistic} Here, we assume that the error rate per unit
 length varies smoothly along the channel. For example, the errors
for different parts of the channel are almost independent.
\bibitem{Peres} The decomposition of the quantum state
into the tensor product of the logical qubits and ancillary qubits is
a mathematical one. In the
actual physical system, the state of
the logical qubits is delocalized among all physical qubits.
Such a delocalization is necessary for both error correction
and fault-tolerant computation.
See, for example, A. Peres, Los Alamos e-Print archive 
(available at http://xxx.lanl.gov/abs/quant-ph/9609015).
\bibitem{scaling} Efficient quantum error correcting
schemes exist for reducing the error rate to an exponentially small amount
(see discussion, available at  http://xxx.lanl.gov/abs/quant-ph/984035.shl)
[that is, Supplementary Note 3 in this reprint version]. 



\bibitem{Newnote} Such an ``$(N-m)$-singlets-or-not'' measurement
can be performed if Alice and Bob bring the two halves
of each EPR pair together to perform a measurement
along a Bell basis. This is a very subtle point because such a Bell measurement
is not actually performed and, indeed, could not be performed without
bringing the two halves together. Successful cheating thus means that the
actual verification test is passed, but a hypothetical second test of bringing
the remaining pairs back into the same laboratory and
measuring them in Bell basis would
fail (that is, some of the remaining $N-m$ pairs are shown to be nonsinglets
upon Bell measurements).



\bibitem{Note2} This is a surprising result because Bell basis vectors
are highly entangled and yet
only local operations and classical communication
are allowed here.
The local operations needed are simply single-qubit operations
and bilocal exclusive OR. More specifically,
the three types of operations used are (i) unilateral rotations by
$\pi$ rad, corresponding to $\sigma_x$, $\sigma_y$ and $\sigma_z$;
(ii) bilateral rotations by $\pi/2$ rad; and (iii) bilateral application of
the two-bit quantum exclusive OR (or controlled NOT). These basic operations
plus local measurements and classical communication allow Alice and Bob
to correct quantum errors using the one-way random-hashing scheme by BDSW.
See (18) for details.

\bibitem{appendix} We can safely use
classical probability theory to derive
an explicit bound on Eve's information for any
eavesdropping strategy that passes the verification
test with a probability at least $2^{-r}$ for some parameter $r> 0$.
See also (17).
Here we work in the approximation of reliable local quantum operations
by Alice and Bob. However, this assumption can be relaxed without
changing our essential conclusion.
If all the original
$N$ pairs are singlets, the remaining $N-m$ pairs must
be singlets. Instead of computing the fidelity for the remaining $N-m$ pairs
to be $N-m$ singlets, let us compute the fidelity for the original
$N$ pairs to be $N$ singlets. This will give us a good enough bound on
the fidelity.
With any cheating strategy against the quantum verification scheme by Eve,
let $p_1$ be the total probability for
the state of the $N$ pairs to be $N$ singlets
under the measurement along the Bell basis.
The case of $N$ singlets, which happens with a probability
$p_1$, will automatically pass the verification test.
This case is perfectly fine and secure.
What about the other case? Upon a random-hashing verification scheme that sacrifices
$m$ pairs, the other case (which happens with probability $1- p_1$) will
pass a $m$-round random-hashing verification test with a conditional probability of, at most $2^{-m}$.
Therefore, the probability that a strategy passes
the verification test is given by
\begin{equation}
P({\rm passing}) \leq  p_1 + 2^{-m} ( 1 -p_1) \leq p_1+   2^{-m}.
\label{passing1}
\end{equation}
Eve would be most interested in a cheating strategy that passes the
test with a nonnegligible probability( say at least $2^{-r}$ where
we assume that $ 0 < r \ll m$).
Therefore, we demand that the probability
\begin{equation}
P({\rm passing})  \geq 2^{-r} .
\label{hell}
\end{equation}
Combining Eq.\ (\ref{passing1}) and (\ref{hell}), we find
that
\begin{eqnarray}
p_1+  2^{-m} &\geq&  2^{-r} \nonumber \\
p_1  &\geq&  2^{-r} [ 1 - 2^{-(m-r)}] .
\label{conp1}
\end{eqnarray}
Conditional on passing the verification
test, the fidelity of the $N$ pairs as singlets is given by
\begin{equation}
F'\geq { p_1 \over p_1+ 2^{-m}  }  \geq 
{ 2^{-r} [ 1 - 2^{-(m-r)}] \over  2^{-r} [ 1 - 2^{-(m-r)}] +  2^{-m}}
=  [ 1 - 2^{-(m-r)}]
\label{heaven}
\end{equation}
where Eq.\ (\ref{conp1})
and the fact that $p_1 \over p_1+ 2^{-m} $ is an increasing function of
$p_1$ have been used.
By choosing a value of $m$ that is substantially larger than $r$,
the conditional fidelity can be made very close to $1$.
Therefore, given any parameter $r$, one can increase the
conditional fidelity in Eq.\ (\ref{heaven}) by increasing the number
$m$ of random parities computed.
In summary, consider any eavesdropping strategy that passes an $m$-round
random-hashing verification scheme with a probability at least $2^{-r}$
(where $m \gg r > 0$). From
Eq.~(\ref{heaven}, upon
passing the test, the conditional fidelity of the
$N$ pairs as $N$ singlets is at least $1 - 2^{- (m-r)}$. From
(28), this implies that Eve's information is
exponentially small in $m-r$, more precisely, $2^{-c} + O (2^{-2(m-r)})$,
where $c= m-r  - \log_2 [2 (N-m) + m -r + { 1 \over \log_e 2}]$.


\bibitem{Semi} R. B. Griffiths and C.-S. Niu, {\it\prl} {\bf 76},
 3228 (1996).
\bibitem{notenew}
Our classical argument applies to the
$N$-Bell basis, whose basis vectors are highly entangled.
It is perhaps surprising at first 
that the coarse-grained probabilities
of a quantum mechanical experiment involving only local operations
and classical communication can have a classical interpretation
with respect to such a highly non-local basis. Put in another way,
given the lesson from the EPR paradox, it is perhaps surprising that classical arguments can still be
used to demonstrate that two distantly separated quantum subsystems are, in fact, highly quantum (that is, highly entangled).


\bibitem{tradeoff}  C. A. Fuchs,  {\it Fortschr.\  Phys.\ } {\bf 46},
535 (1998) and references cited therein.

\bibitem{incident} Incidentally, our result also proves the security of
quantum money proposed by
Wiesner (1).
Indeed, the proof for our second example can be used to derive
a probabilistic bound on the entropy of the combined system consisting
of the
quantum banknote and the bank. 
Consequently, any double-spending strategy
will almost surely fail in the verification step (as in
BB84) done by the bank because
this entropy will no longer be close to zero.
\bibitem{Smolin} J. Smolin, personal communication.
\bibitem{Ack} H.-K. Lo particularly thanks A. Ekert for pressing him
 to investigate the security of QKD. We thank
 numerous colleagues, including
 C. H. Bennett, G. Brassard, I. Chuang, D. P. DiVincenzo, C. A. Fuchs, N.
 Gisin,
 D. Gottesman, E. Knill,
 D. W. C. Leung, N. L\"{u}tkenhaus, D. Mayers, S. Popescu, J. Preskill, 
 J. Smolin, T. Spiller, A. Steane,
 and  A. C.-C. Yao for invaluable conversations and
 suggestions. Many helpful suggestions from an anonymous referee are
 gratefully acknowledged.
 H. F. Chau is supported by Hong Kong Government
 RGC grant HKU~7095/97P.

\end{references}
\end{document}